\documentclass[%
reprint,
superscriptaddress,
amsmath,amssymb,
aps,
pra,
floatfix,
]{revtex4-2}
\usepackage[T1]{fontenc}
\usepackage{lmodern,microtype,amsmath,amssymb,graphicx,flafter}
\usepackage[colorlinks=true,citecolor=blue,urlcolor=blue]{hyperref}
\usepackage{enumitem}
\usepackage[dvipsnames]{xcolor}

\begin{document}
\title{Entanglement Sudden Death: Eberly’s Work and the Limits of Local Dynamics}
\author{Abhaya S. Hegde}
\email{a.hegde@rochester.edu}
\affiliation{Department of Physics and Astronomy, University of Rochester, Rochester, New York 14627, USA}
\affiliation{University of Rochester Center for Coherence and Quantum Science, Rochester, New York 14627, USA}

\author{Songbo Xie}
\email{sxie2@ncsu.edu}
\affiliation{Department of Electrical and Computer Engineering, North Carolina State University, Raleigh, North Carolina 27606, USA}
\date{September 2026}

\begin{abstract}
Entanglement sudden death (ESD), explored through a series of works by Yu and Eberly, supports the notion of entanglement vanishing abruptly and completely at a finite time, even when the underlying open-system dynamics remain smooth. 
In this article, we revisit the development of ESD from bipartite origins to genuine multipartite entanglement.
We first illustrate the canonical mechanism of ESD with two independently damped qubits, for which the concurrence vanishes at a finite time while populations and coherences remain nonzero. 
We next present our results on multipartite entanglement for generalized GHZ states under depolarization. 
Global and independent local depolarizing dynamics can be chosen to produce identical reduced evolution for every individual qubit, yet they yield different genuine-multipartite-entanglement sudden death times, whose ordering can reverse with the initial state. 
For $N$-qubit GHZ states, this distinction becomes asymptotically stronger since the ESD time scales as $1/N$ under independent local depolarization, whereas under global depolarization it approaches a finite value of $\ln(2/\gamma)$ as $N \to \infty$. Echoing a central lesson of Prof.~Eberly's work, we report the lifetime of multipartite entanglement is not fixed by the local dynamics, but rather depends on the evolution of the joint quantum state.
\end{abstract}

\maketitle

\section{Introduction}

Coherence and entanglement are two central signatures of quantum behavior. Coherence describes quantum superposition within a system, whereas entanglement concerns quantum correlations shared between two or more subsystems. While both are vulnerable to environmental noise, the decay of coherence follows a familiar pattern. For a system weakly coupled to its surroundings, coherence typically decreases exponentially in time: after each characteristic decay time, only a fixed fraction remains. This is the familiar half-life picture of decay, in which coherence becomes progressively smaller but reaches zero only asymptotically. What, then, happens to entanglement? If the subsystems that share it are each undergoing ordinary exponential decoherence, should their entanglement decay in the same way and survive for every finite time?

Surprisingly, the answer is no. In a series of early studies, Yu and Eberly showed that entanglement can decay on a time scale very different from that of single-system coherence~\cite{YuEberly2002Phonon,YuEberly2003Dephasing}. The contrast became especially clear for two initially entangled qubits, each undergoing independent spontaneous emission~\cite{YuEberly2004FiniteTime}. The excited-state populations and coherences of the individual qubits decay continuously, as expected, yet their entanglement can reach exactly zero after a finite time. Yu and Eberly referred to this finite-time loss of entanglement as \emph{entanglement sudden death} (ESD). Thus, even when each qubit follows the usual asymptotic decay, the entanglement shared between them need not do so.

It became clear over the years that ESD was not limited to spontaneous emission processes. Finite-time disentanglement was found to occur under not only quantum noise but also classical noise~\cite{YuEberly2006ClassicalNoise}.
This phenomena were also observed in cavity-QED systems, where entanglement can disappear for a finite interval and later revive~\cite{YonacYuEberly2006Atoms}. 
Subsequent studies expanded the focus for initial states from pure states to broad families of mixed states, and concluded that the occurrence and timing of ESD depend strongly on the specific choice of initial state~\cite{YuEberly2007XStates,QianEberly2012Initial}.
The effect was subsequently observed in an all-optical experiment~\cite{AlmeidaEtAl2007Experiment}.
These results established ESD as a general feature of entanglement dynamics in open quantum systems, offering a view that entanglement loss cannot be understood simply as a faster form decoherence~\cite{YuEberly2006Bipartite,YuEberly2009Review,aolita2015open}.

The same question on the loss of entanglement in finite-time acquires even more subtlety in multipartite systems.
With three or more parties, the genuine multipartite entanglement can vanish while entanglement still survives among smaller subsets of the overall system. 
In this case, the state has become biseparable rather than fully separable. 
Multipartite ESD therefore asks not only whether entanglement disappears, but which form of entanglement is lost first. Eberly's later work addressed this problem directly, including the derivation of an exact genuine multipartite concurrence for $N$-qubit $X$ states and studies of the decay and robustness of genuine multipartite entanglement under spontaneous emission~\cite{hashemirafsanjaniGenuinelyMultipartiteConcurrence2012,XieYounisEberly2023Robustness}.

This article is written in memory of the late Professor Joseph H. Eberly, whose work with his collaborators opened a new way of thinking about the lifetime of entanglement in open quantum systems.
We begin in Sec.~II with a brief historical account of results on ESD and the questions that grew out of Eberly's pioneering studies.
In Sec.~III, we use a simple bipartite example to show explicitly how entanglement can vanish at a finite time while the density matrix continues to evolve smoothly. 
In Sec.~IV, we extend this analysis to genuine tripartite entanglement by comparing entanglement dynamics under local and global sources of depolarization channels. The two noise models are constructed to give the same reduced dynamics for each individual qubit, but they drive the full system differently. 
We find that the lifetime of genuine multipartite entanglement is not fixed by the local dynamics: two noise processes that are indistinguishable at the single-qubit level can produce different ESD times, and their ordering can even reverse depending on the initial state.

The motivation for this article is to commemorate the life and work of Prof.~Eberly by expanding upon their final research avenues. 
As a Ph.D. student in Prof.~Gabriel Landi's group at the University of Rochester, Abhaya Hegde interacted closely with Eberly, beginning with the introductory quantum optics course \texttt{PHYS531} in Fall 2022.
The course ended with a requirement to turn in a term paper whose central result is developed further in the present article.

Songbo Xie was Prof.~Eberly's final Ph.D. student, graduating from Rochester in May 2024 after working with him for nearly six years. During this time, they collaborated closely on the quantification of genuine multipartite entanglement, including its geometric characterization, experimental estimation, dynamics, and sudden death. These projects formed an important part of Songbo's doctoral research and scientific training under Prof.~Eberly. In particular, their work on entanglement sudden death provides a direct scientific connection to the subject developed in the present article. The two authors were brought together by their shared connection to Prof.~Eberly and their common interest in the physics of ESD,  which eventually led to this article.

\section{Eberly and the Development of Entanglement Sudden Death}

Eberly's work on entanglement sudden death grew from a simple question about lifetime: should entanglement decay under environmental noise in the same gradual way as coherence?
With his pioneering studies in early 2000s that established the complete loss of entanglement in finite time, 
the line of enquiry shifted to what determines its death time, why some states undergo ESD while others do not, and under what conditions the effect be avoided or even reversed. 
Later on, as the focus moved from bipartite to multipartite systems, Eberly found an additional issue needing to be addressed: when several kinds of entanglement can coexist, what exactly does it mean for entanglement to die?
We review each phase of his research program on ESD in this section.

\subsection{Discovery: entanglement has its own lifetime}

The possibility that entanglement could behave differently from conventional measures of coherence had appeared before the phenomenon acquired the name ESD.
An early hint appeared in the work of \.{Z}yczkowski et al., who studied the dynamics of bipartite entanglement under repeated noisy quantum channels in 2001~\cite{Zyczkowski2001}. 
In the presence of competing unitary and dissipative dynamics, they found intervals over which the entanglement vanished completely followed by revivals, referring to the resulting regions of zero entanglement as \emph{entanglement plateaus}.

Against this background, Yu and Eberly in 2002 studied two atoms subject to phonon-induced pure dephasing as a way to distinguish the dynamics of entanglement from those of local coherence.
They found that initially entangled states could respond very differently to the same dephasing process. While some were relatively robust, the others were fragile and lost entanglement much faster than the individual qubits lost coherence. As they summarized, \emph{``there exist two different decoherence time scales---entanglement decay time and local dephasing time.''} In 2003, they extended this comparison to local and global phase noise~\cite{YuEberly2003Dephasing}, finding again that the time scale for entanglement decay could be significantly shorter than the single-qubit dephasing time. 
These results showed that the persistence of bipartite entanglement could not, in general, be inferred from the coherence time of the individual subsystems.

Meanwhile, related studies showed that this distinction from coherence decay could be made sharper. 
Under suitable open-system dynamics, complete disentanglement was found to occur after a finite time. 
Di\'osi showed in 2003 that a depolarizing channel acting on a two-state system becomes entanglement breaking after a finite interval, thereby eliminating its entanglement with any otherwise noninteracting external system~\cite{Diosi2003}. 
Dodd and Halliwell subsequently obtained a corresponding result for continuous-variable open systems, proving for a class of two-particle master equations that arbitrary initial states attain separability after finite time~\cite{Dodd2004}. 
Thus, the finite-time separability was found to be applicable across different models and various measures of entanglement.

The distinctive contribution of Yu and Eberly arrived in 2004, when they demonstrated finite-time disentanglement in perhaps its most elementary open-system setting: two initially entangled two-level atoms undergoing spontaneous emission into vacuum reservoirs~\cite{YuEberly2004FiniteTime}. For suitable initial states, the excited-state populations and coherences of the atoms exhibited the usual asymptotic relaxation, but the Wootters concurrence~\cite{Wootters1998Concurrence} was shown to reach exactly zero at a finite time. 
In their words, \emph{``complete disentanglement can be reached after only a finite-time, whereas more familiar local decoherence processes take an infinite time to be complete''}. 
This disappearance of entanglement was later termed as \emph{entanglement sudden death}.
Although this phenomena had appeared in earlier studies, the work of Yu and Eberly established its significance in the familiar setting of two-qubit decoherence and helped initiate its systematic study as a distinct feature of entanglement dynamics.

Importantly, the term ``sudden'' here refers only to the finite time at which entanglement vanishes. In other words, the density matrix itself continues to evolve smoothly throughout and beyond this marked time. In geometric terms, the evolving state crosses the boundary of the separable set at a finite time, even while its populations and coherences remain nonzero. ESD is therefore a finite-time loss of entanglement between the subsystems, and does not confer a meaning of sudden change in their open-system dynamics.

\subsection{Understanding and controlling entanglement sudden death}

After the 2004 result, an immediate question was how general this phenomenon of finite-time disentanglement could be.
Yu and Eberly addressed this problem through a broad family of two-qubit mixed states, the so called $X$-states, first studied in a 2005 preprint which was later published in 2007~\cite{YuEberly2007XStates}. 
As they noted, the $X$-form \emph{``arises naturally in a wide variety of physical situations''} and includes Bell and Werner states as special cases. 
In the computational product basis ${|00\rangle,|01\rangle,|10\rangle,|11\rangle}$,
\begin{equation}
    \rho_X=
    \begin{pmatrix}
        a&0&0&w\\
        0&b&z&0\\
        0&z^*&c&0\\
        w^*&0&0&d
    \end{pmatrix},
    \quad a+b+c+d=1,
\end{equation}
the concurrence is
\begin{equation}
    C(\rho_X)=2\max\bigl\{0,\,|w|-\sqrt{bc},\,|z|-\sqrt{ad}\bigr\}.
\label{eq:eberly-x-concurrence}
\end{equation}
This expression provided a direct picture of how ESD occurs as a function of coherence and populations. 
Entanglement is present only if at least one of the two quantities $|w|-\sqrt{bc}$ or $|z|-\sqrt{ad}$ is positive. Under dissipative evolution, the populations and coherences change continuously, but both quantities can become nonpositive at a finite time, at which point the concurrence vanishes exactly. 

The same analysis also showed that entanglement survival depends strongly on the initial state. Importantly, a local unitary transformation can leave the initial concurrence unchanged while changing the subsequent lifetime of the entanglement, and in some cases avoids ESD altogether. Yu and Eberly concluded that \emph{``ESD may be manipulated by preparatory transformation that is purely local.''} The question had therefore moved beyond whether ESD occurs \emph{at all} to whether its occurrence could be controlled before the noisy evolution begins.

The next insightful discovery pertained to whether finite-time disentanglement was specific to spontaneous emission and the vacuum reservoir considered in the celebrated example from 2004. Yu and Eberly showed in 2006 that classical phase noise alone can produce ESD without energy relaxation~\cite{YuEberly2006ClassicalNoise}, describing the effect as \emph{``a new kind of nonlocal decoherence.''} 
Around the same time, Y{\"o}na{\c c}, Yu, and Eberly found finite-time loss of atomic entanglement for two atoms interacting separately with their own Jaynes--Cummings cavities~\cite{YonacYuEberly2006Atoms}. Yu and Eberly then placed these results in a broader open-system framework~\cite{YuEberly2006Bipartite}. Although decay rates are additive for the individual systems, a composite system \emph{``need not and generally does not exhibit relaxation-rate additivity even though the separate parts do.''} ESD was therefore not tied to spontaneous emission or to a particular reservoir model, but reflected a more general distinction between local decoherence and bipartite entanglement dynamics.

The $X$-state formula in Eq.~\eqref{eq:eberly-x-concurrence} also raised a technical question: was the sudden disappearance of entanglement merely a consequence of the mathematical form of concurrence? Yu and Eberly addressed this by extracting its variable part inside the maximum in Eq.~\eqref{eq:eberly-x-concurrence}, and using it as a measure of \emph{`` `distance,' either positive or negative, from the separability boundary''}~\cite{YuEberly2007Negative}. In this picture, the zero of concurrence marks the point at which a continuously evolving state reaches that boundary. Importantly, \emph{``the boundary of separability of a state is independent of the entanglement measure.''} The density matrix itself remains smooth and the suddenness of ESD comes from crossing the separability boundary at a finite time.

In 2007, Almeida and collaborators used an all-optical setup to demonstrate the difference between local, single-particle dynamics and the global dynamics of an entangled two-qubit system coupled to independent environments~\cite{AlmeidaEtAl2007Experiment}. While the environment-induced decay of each individual system remained asymptotic, the entanglement disappeared completely after a finite interval. In an accompanying \emph{Science} Perspective, Eberly and Yu described the experiment as an \emph{``elegantly clean way to check and to confirm the existence''} of ESD, which they characterized as \emph{``novel among known relaxation effects''}~\cite{EberlyYu2007End}. With this, ESD had thus progressed from a theoretical prediction to an experimentally observed effect.

In parallel, Eberly's research also focused on this aspect of entanglement decay in cavity-QED systems. 
These studies helped answer a different question: does the sudden death of entanglement between a particular pair imply that the entanglement has been irreversibly lost? In a four-partite atom--field model, Y{\"o}na{\c c}, Yu, and Eberly found that the loss of atom--atom entanglement could be \emph{``instantly compensated by concurrence gain in $C^{ab}$''} for the two cavity fields, revealing entanglement transfer within the larger system~\cite{YonacYuEberly2007Pairwise}. Y{\"o}na{\c c} and Eberly later found collapse and revival of entanglement between remote qubits driven by independent quantized optical fields~\cite{YonacEberly2008Remote}. These results showed that pairwise entanglement can disappear momentarily either through transfer or by recovering later, rather than being irreversibly lost.

By 2009, the accumulated results had established ESD as a distinct form of entanglement decay. In their \emph{Science} article ``\textit{Sudden Death of Entanglement},'' Yu and Eberly reviewed the development of the subject and described ESD as a \emph{``new route for dissipation''}, distinct from the familiar asymptotic behavior of ordinary relaxation~\cite{YuEberly2009Review}. Attention then began to turn from identifying ESD to controlling it. Y{\"o}na{\c c} and Eberly showed that suitably chosen quantum noise could preserve joint coherence and enable the hiding and recovery of entanglement~\cite{yonac2009reverse}. They also studied coherent-state control of remote, noninteracting qubits and the resulting collapse and revival of their entanglement~\cite{yonac2010coherent}.

The role of the environment and the initial state then became part of this control problem. Yu and Eberly studied two spins under classical Ornstein--Uhlenbeck noise, obtaining exact entanglement dynamics beyond the Markovian limit~\cite{YuEberly2010NonMarkovian}. Qian and Eberly also approached the problem from the viewpoint of state preparation~\cite{QianEberly2012Initial}. For a class of Bell-superposition states, the occurrence and timing of ESD \emph{``depend on only three parameters: initial purity, entanglement and excitation.''} This allowed them to identify ESD-free and ESD-susceptible regions of the initial-state space and determine the onset time within the susceptible region. The susceptibility to ESD could therefore be anticipated, and in some cases avoided or delayed.

The effort to delay or prevent ESD also led to alternative ways to improve quantum error correction. In 2014, Y{\"o}na{\c c} and Eberly studied amplitude damping, pure phase damping, and combinations of the two while following both fidelity and entanglement~\cite{yonac2014effects}. Their central finding was that \emph{``error correction affects entanglement and fidelity differently''}: for some initial states, error correction increases fidelity while decreasing entanglement, or vice versa. Fidelity alone therefore was inadequate to determine how well entanglement had been protected.
The subfield of entanglement dynamics had matured by then to explore many avenues of ESD in bipartite systems.

\subsection{From bipartite to multipartite entanglement}
Moving beyond two parties into multipartite systems complicates the notion of entanglement death. 
Asking when entanglement dies first then requires specifying which form of entanglement is being considered.
In a bipartite system, the loss of entanglement leaves a separable state. In a multipartite system, genuine multipartite entanglement can vanish while entanglement still survives among smaller groups. An important result emerged in 2012, when Hashemi Rafsanjani, Huber, Broadbent, and Eberly obtained an exact concurrence for $N$-qubit $X$ states~\cite{hashemirafsanjaniGenuinelyMultipartiteConcurrence2012}. They then applied this formula to a family of generalized $N$-qubit GHZ initial states undergoing local amplitude damping. Within this family, \emph{``only one type of GHZ state loses its entanglement in finite time''}, whereas for the others the $N$-partite entanglement decays asymptotically. This extended the study of ESD from pairwise entanglement to genuine multipartite entanglement for GHZ-type initial states.

Hashemi Rafsanjani and Eberly later studied coherent control of multipartite entanglement for an arbitrary number of remote qubits initially prepared in a symmetric GHZ state~\cite{rafsanjani2015coherent}. They compared all-party entanglement with weak inseparability and found different dynamics. For $N\geq4$, the all-party entanglement does not revive after the initial collapse, while \emph{``weak inseparability repeatedly revives from zero to a substantial non-zero value''} even for large $N$. The same coherent evolution can therefore have very different effects on different levels of multipartite entanglement.

A different type of sudden change in entanglement behavior appeared in 2021. Ding, Xie, and Eberly studied a three-partite atom--field system undergoing fully coherent pure-state evolution and found \emph{``nonanalytic slope discontinuities in the time record''} of entanglement~\cite{ding2021sudden}. The entanglement sharing could suddenly freeze at a constant value, later thaw, or remain permanently frozen, without any decohering environment. Importantly, the sudden changes in entanglement dynamics need not originate from dissipation, but can also arise from constraints on how entanglement is shared among several parties.

Later in 2024, about two decades after the 2004 spontaneous-emission result, Xie, Younis, and Eberly returned to the original ESD problem, asking how spontaneous emission affects genuine tripartite entanglement~\cite{XieYounisEberly2023Robustness}. They examined generalized GHZ states, $W$-type states, and more general multiparitite superpositions. 
They identified the fully excited component $|11\cdots1\rangle$ as the \emph{``unique `ESD-trigger state' ''}. 
This observation suggested that increasing the number of qubits and spontaneous-emission channels does not necessarily make genuine multipartite entanglement more vulnerable to ESD. 
As the number of qubits increases, the Hilbert space grows while this single trigger state occupies a progressively smaller fraction of it, thereby increasing robustness of genuine multipartite entanglement against sudden death.

A geometric approach to genuine multipartite entanglement became another part of Eberly's later work. 
Particularly, it addressed the lack of a measure that identified genuine multipartite correlations without which the proposed extension of sudden death to multipartite systems is meaningless.
Before asking when genuine multipartite entanglement disappears, one must have a measure whose vanishing correctly identifies the biseparable boundary.
In 2021, Xie and Eberly introduced the concurrence fill, a genuine tripartite entanglement measure constructed from the area of a concurrence triangle~\cite{xie2021triangle}. This geometric framework was later developed in a broader study of multipartite entanglement~\cite{xie2024multipartite}.

Across more than two decades, Eberly's work transformed ESD from a surprising two-qubit prediction into a broader understanding of entanglement dynamics. His studies showed that entanglement can have its own lifetime, can disappear or return while local evolution remains smooth, and can take qualitatively different forms in multipartite systems. This body of work made ESD a lasting part of the modern study of open quantum systems.

\section{A simple canonical example of Entanglement Sudden Death}
To make the mechanism of sudden death explicit, we begin with a simple toy model that shows how it arises during a typical smooth dissipative dynamics. Following Yu and Eberly's foundational work on spontaneous emission~\cite{YuEberly2004FiniteTime}, we consider two noninteracting two-level atoms, each radiating into its own vacuum reservoir. Let the atoms be initially prepared in the state
\begin{equation}
|\psi(0)\rangle=\frac{1}{\sqrt5}|00\rangle+\frac{2}{\sqrt5}|11\rangle,
\end{equation}
where $|0\rangle$ and $|1\rangle$ denote the ground and excited levels. The reservoirs are independent, and there is no subsequent interaction between the two atoms.



We model the spontaneous emission of each atom by the standard amplitude-damping channel. For a single atom, the channel can be written as
\begin{equation}
\mathcal{E}_{t}(\rho)=K_{0}(t)\rho K_{0}^{\dagger}(t)+K_{1}(t)\rho K_{1}^{\dagger}(t),
\end{equation}
with
\begin{equation}
\begin{split}
    K_{0}(t)&=|0\rangle\langle0|+\sqrt{q(t)}\,|1\rangle\langle1|,\\
    K_{1}(t)&=\sqrt{1-q(t)}\,|0\rangle\langle1|,\\
    \text{with\ }  q(t)&=e^{-\gamma t}.
\end{split}
\end{equation}
Here $\gamma$ is the spontaneous-emission rate and $q(t)$ is the survival probability of an excitation initially in the excited state. Since the two atoms radiate into independent reservoirs, their joint evolution is simply
\begin{equation}
\rho(t)=\left(\mathcal{E}_{t}\otimes\mathcal{E}_{t}\right)\rho(0),
\qquad
\rho(0)=|\psi(0)\rangle\langle\psi(0)|.
\end{equation}

Applying this channel to the initial state gives, in the basis
$\{|00\rangle,|01\rangle,|10\rangle,|11\rangle\}$,
\begin{equation}
\rho(t)=
\begin{pmatrix}
\frac{1}{5}+\frac{4}{5}(1-q)^{2} & 0 & 0 & \frac{2}{5}q\\
0 & \frac{4}{5}q(1-q) & 0 & 0\\
0 & 0 & \frac{4}{5}q(1-q) & 0\\
\frac{2}{5}q & 0 & 0 & \frac{4}{5}q^{2}
\end{pmatrix}.
\end{equation}
All elements of the density matrix evolve smoothly in time. We now ask whether the entanglement follows the same behavior.

\begin{figure}[t]
    \centering
    \includegraphics[width=\linewidth]{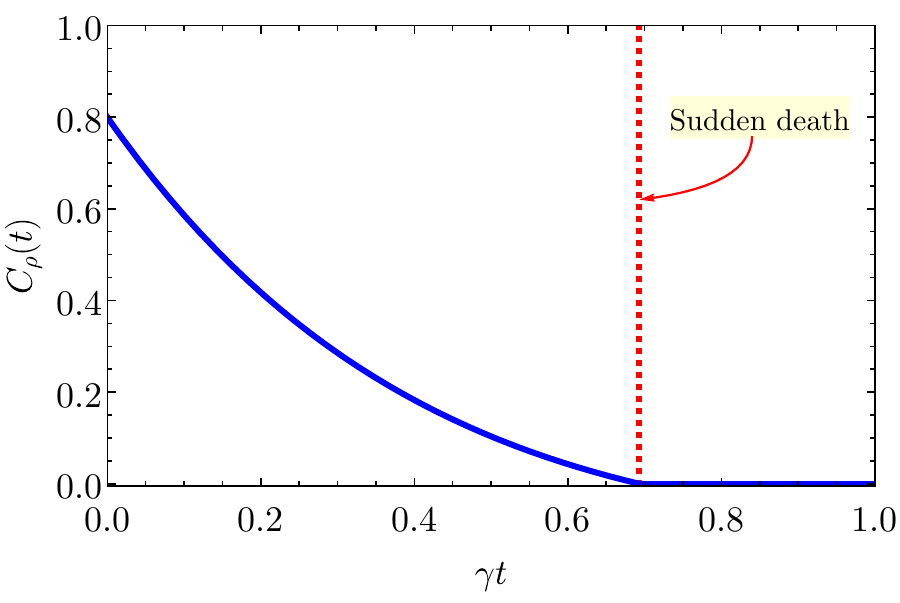}
    \caption{Concurrence from Eq.~\eqref{eq:concurrence_example}. The dashed red vertical line marks the time at which entanglement vanishes abruptly and completely.}
    \label{fig:concurrence_example}
\end{figure}

To determine when the atoms remain entangled, we use Wootters' concurrence \cite{Wootters1998Concurrence}. For a two-qubit state $\rho$, it is defined as
\begin{equation}
C_\rho(t)=\max\bigl\{0,\lambda_{1}(t)-\lambda_{2}(t)-\lambda_{3}(t)-\lambda_{4}(t)\bigr\},
\end{equation}
where $\lambda_{i}(t)$ are the square roots of the eigenvalues of
$\rho\tilde{\rho}$ in decreasing order, with
\begin{equation}
\tilde{\rho}=(\sigma_{y}\otimes\sigma_{y})\rho^{*}
(\sigma_{y}\otimes\sigma_{y}).
\end{equation}
For the $X$-state above, the concurrence reduces to
\begin{equation}
C_\rho(t)=2\max\biggl\{
0,\,
|\rho_{00,11}(t)|-\sqrt{\rho_{01,01}(t)\rho_{10,10}(t)}
\biggr\}.
\end{equation}
Substituting the matrix elements yields
\begin{equation}
\label{eq:concurrence_example}
C_\rho(t)=
\frac{4}{5}q\,\max\{0,2q-1\}.
\end{equation}
The initial concurrence is therefore $C_{\rho}(0)=4/5$. As the atoms emit, the concurrence decreases continuously until the two contributions inside the maximum become equal. This occurs at
\begin{equation}
q(t_{*})=\frac{1}{2},
\end{equation}
and hence
\begin{equation}
t_{*}=\frac{\ln2}{\gamma}
\end{equation}
denotes the time of concurrence vanishing completely.
For all later times, $q<1/2$ and the concurrence remains exactly zero, as shown in Fig.~\ref{fig:concurrence_example}.

This is the essential mechanism of entanglement sudden death. At $t=t_{*}$, neither the spontaneous-emission process nor the underlying density-matrix evolution has terminated. In fact, the excited-state population of either atom at $t=t_*$ is $2/5$,
and the coherence being $\rho_{00,11}(t_{*})=1/5$, both remaining nonzero. The entanglement nevertheless vanishes because the surviving coherence is no longer large enough, relative to the populations generated in $|01\rangle$ and $|10\rangle$, to keep the state outside the separable set.
Most importantly, this vanishing of correlations is not due to the specific choice of concurrence as the entanglement metric, as it vanishes universally at $t_*$ across all measures of entanglement~\cite{YuEberly2007Negative}.

\section{Genuine multipartite entanglement sudden death under depolarization}

Our result addresses the following question: if the reduced dynamics of each qubit is held fixed, can the spatial organization of the noise still affect the lifetime of genuine multipartite entanglement? To this end, we compare the evolution of generalized GHZ states under independent local depolarization and global depolarization.

Differences in entanglement structure under local and global dynamics have previously been investigated in Ref.~\cite{Filippov2013}. 
While Sk and Panigrahi~\cite{Panigrahi2022} showed that correlations between successive uses of noisy channels can suppress or even prevent ESD in GHZ-class states, our analysis pertains to a different aspect of ESD. 
We show that the lifetime of genuine multipartite entanglement is not fixed by the local dynamics alone, since two noise processes that are indistinguishable at the single-qubit level can yield different ESD times, with their ordering even reversing depending on the initial state.
We also find qualitatively different system-size scaling of entanglement lifetime under local and global depolarizing dynamics.

The reference state is~\cite{Panigrahi2022, XieYounisEberly2023Robustness},
\begin{equation}
|\psi_\theta\rangle =\cos\theta\,|000\rangle+\sin\theta\,|111\rangle,
\qquad 0 < \theta \leq \frac{\pi}{4},
\end{equation}
where the parameter $\theta$ determines the initial entanglement. A maximally entangled GHZ state results at $\theta = \pi/4$, while a product state is approached as $\theta \to 0$.
We first subject this state to a global white noise that affects the joint state and then compare it with independent local depolarization of the three qubits.

\begin{figure*}[t]
    \centering
    \includegraphics[width=\linewidth]{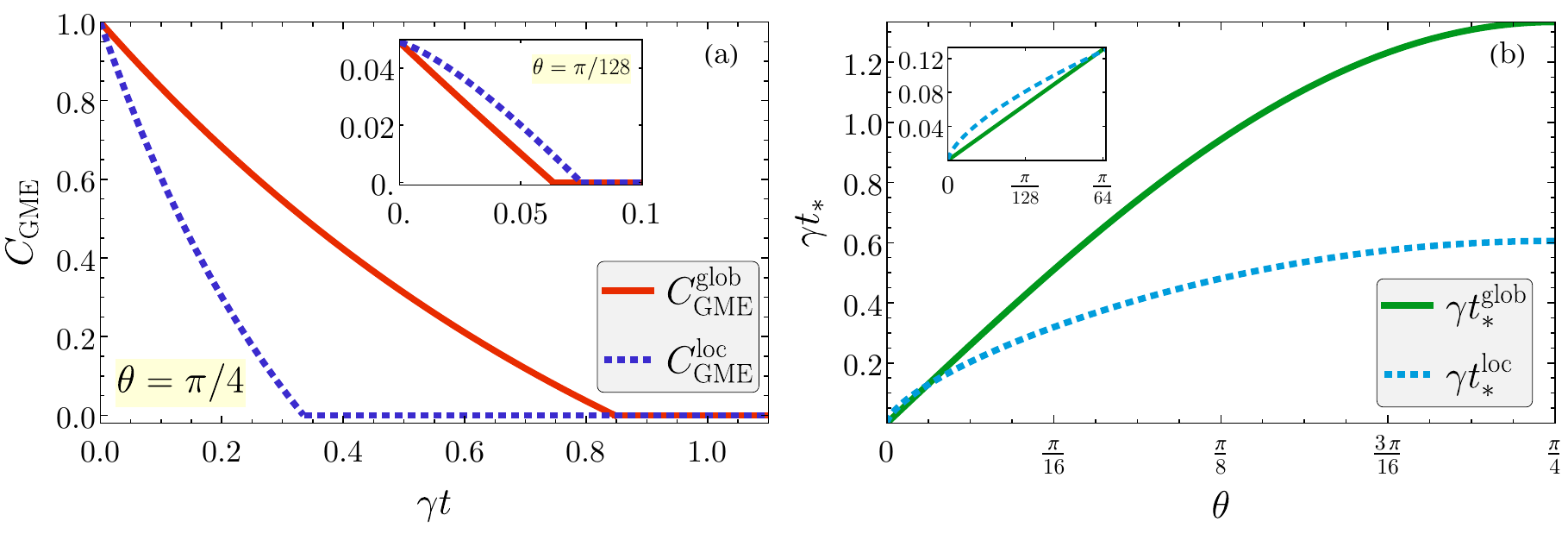}
    \caption{(a) Concurrence as a function of time for GHZ state initialized in maximally entangled state. The solid red line refers to dynamics under global depolarization and the dotted blue represents qubits subjected to independent depolarization. While independently depolarized states lose tripartite genuine entanglement faster than global white noise for this initial state, for sufficiently weakly entangled states, such as $\theta = \pi/128$, have opposite order in which dynamics retains genuine correlations longer, as shown inset. (b) Entanglement sudden death time as a function amount of initial entanglement parametrized by the superposition angle $\theta$ in the initial GHZ state.}
    \label{fig:GMEdepolarizing}
\end{figure*}

\subsection{Global depolarization of GHZ state}
We consider a Pauli channel under which the three-qubit GHZ state evolves as
\begin{equation}
\label{eq:rho_global}
\rho_{\theta}(t) = p(t)|\psi_\theta\rangle\langle\psi_\theta| + [1-p(t)]\frac{\mathbb{I}_8}{8},
\quad
p(t)=e^{-\gamma t}.
\end{equation}
where $\gamma$ denotes the depolarization rate. This dynamics drives the three-qubit state to the maximally mixed state $\mathbb{I}_8/8$ and we therefore refer to it as global depolarization.
The only nonzero off-diagonal elements in the density matrix turn out to be
\begin{equation}
\label{eq:GHZ_coherence}
\rho_{000,111}
=\rho_{111,000}
=p\cos\theta\sin\theta.
\end{equation}
It is easy to see that the six populations excluding $|000\rangle$ and $|111\rangle$ are all equal to $(1-p)/8$, rendering an $X$-state for which the genuine multipartite concurrence is known exactly~\cite{hashemirafsanjaniGenuinelyMultipartiteConcurrence2012}.

For the chosen initial state, the concurrence quantifying genuine multipartite entanglement reduces to
\begin{equation}
C^{\rm glob}_{\rm GME}(t, \theta)
=
\max\biggl\{0,\,
p(t)\sin(2\theta)-\frac{3}{4}\bigl[1-p(t)\bigr]
\biggr\}.
\end{equation}
The two terms in the non-zero concurrence have a direct interpretation: the first comes from the surviving GHZ coherence terms, while the second refers to the populations introduced by depolarization. Thus, the genuine tripartite entanglement survives only while the coherent contribution exceeds this population contribution.

Setting the expression inside the maximum to zero gives
\begin{equation}
\label{eq:pglob}
p_*^{\rm glob}=\frac{3}{3+4\sin(2\theta)},
\end{equation}
and therefore
\begin{equation}
t_*^{\rm glob}=\frac{1}{\gamma}
\ln\biggl[1+\frac{4}{3}\sin(2\theta)\biggr].
\end{equation}
For the range considered here, $0<\theta\le\pi/4$, we have $\sin(2\theta)>0$, so that $0<p_*^{\rm glob}<1$ and $t_*^{\rm glob}>0$. Thus, every initially entangled state in this family undergoes genuine multipartite entanglement sudden death at a finite time.

Importantly, the GHZ coherence in Eq.~\eqref{eq:GHZ_coherence} remains nonzero at $t_*^{\rm glob}$ and, in fact, at every finite time. The complete loss of genuine tripartite entanglement therefore does not require the coherence itself to vanish. Instead, ESD occurs when the surviving coherence becomes too small relative to the populations generated by depolarization to maintain genuine tripartite entanglement.


For the maximally entangled GHZ state, $\theta=\pi/4$, the concurrence simplifies to
\begin{equation}
\label{glob_conc}
C_{\rm GME}^{\rm glob}(t,\pi/4)
=
\max\biggl\{0,\frac{7p(t)-3}{4}\biggr\}.
\end{equation}
The corresponding ESD time is
\begin{equation}
\gamma t_*^{\rm glob}=\ln(7/3)\approx0.847.
\end{equation}
This is the longest ESD time within the family $0<\theta\le\pi/4$. More generally, $t_*^{\rm glob}$ increases monotonically with $\theta$, so states with weaker initial entanglement lose their genuine tripartite entanglement earlier. In the limit $\theta\to0$, the initial state approaches the product state $|000\rangle$, and $t_*^{\rm glob}\to0$ accordingly.

The disappearance of genuine tripartite entanglement at $t_*^{\rm glob}$ does not necessarily imply that all entanglement has been lost. To examine whether entanglement still survives across a one-versus-two bipartition, we consider the partial transpose with respect to the first qubit and define the negativity as $\mathcal{N}_{1|23}=(\|\rho^{T_1}\|_1-1)/2$ \cite{vidal2002computable}. For the state in Eq.~\eqref{eq:rho_global}, we obtain
\begin{equation}
\mathcal{N}_{1|23}=\max\biggl\{0,\frac{p[1+4\sin(2\theta)]-1}{8}\biggr\}.
\end{equation}
By permutation symmetry, the same result holds for the other two one-versus-two bipartitions. The negativity therefore remains nonzero until
\begin{equation}
\gamma t=\ln\bigl[1+4\sin(2\theta)\bigr].
\end{equation}

This threshold occurs later than the disappearance of genuine tripartite entanglement,
\begin{equation}
\gamma t_*^{\rm glob}=\ln\bigl[1+\tfrac{4}{3}\sin(2\theta)\bigr].
\end{equation}
Thus, for
\begin{equation}
\ln\bigl[1+\tfrac{4}{3}\sin(2\theta)\bigr]\leq\gamma t<\ln\bigl[1+4\sin(2\theta)\bigr],
\end{equation}
the genuine tripartite entanglement has already vanished, while the state remains entangled across every one-versus-two bipartition.

Notice that each one-versus-two partition is a $2\times4$ bipartite system. Unlike the $2\times2$ and $2\times3$ cases, positivity under partial transpose is not sufficient to establish separability in $2\times4$ \cite{peres1996separability,HORODECKI19961}. Therefore, a nonzero negativity unambiguously certifies entanglement across the corresponding bipartition, whereas vanishing negativity does not rule out positive partial transpose (PPT) entanglement. Consequently, for $\gamma t\geq\ln[1+4\sin(2\theta)]$, we can conclude only that the state is PPT across all three one-versus-two bipartitions, not that it is necessarily biseparable.

\subsection{Local depolarization of GHZ state}
A useful comparison for the previous scenario is obtained by applying the single-qubit depolarizing map
\begin{equation}
\label{eq:single_qubit_depol}
\mathcal L_p(\rho)=p\rho+(1-p)\frac{\mathbb{I}_2}{2}
\end{equation}
independently to each qubit:
\begin{equation}
\rho_{\mathrm{loc}}(t)
=\mathcal L_p^{\otimes3}
\bigl(|\psi_\theta\rangle\langle\psi_\theta|\bigr).
\end{equation}
We choose the same depolarization factor $p(t) = e^{-\gamma t}$ as in the global case. With this choice, both global and local depolarization models give exactly the same reduced state for each individual qubit,
\begin{equation}
\rho_j(t)
=p(t) \rho_j(0) + [1-p(t)]\frac{\mathbb{I}}{2}, \quad j \in \{1, 2, 3\}.
\end{equation}

Nevertheless, the joint states evolve differently since, under independent depolarization, the GHZ coherence acquires a factor of $p$ from each qubit and thus decays overall as $p^3$.
Applying the same $X$-state concurrence formula gives
\begin{align}
\label{local_conc}
C_{\mathrm{GME}}^{\mathrm{loc}}(t, \theta)
=
\max\Bigl\{0,\, p^3\sin(2\theta)-&\frac{3}{4}(1-p^2) \Bigr. \\
&\ \Bigl.\times\sqrt{1-p^2\cos^{2}(2\theta)}
\Bigr\} \notag.
\end{align}
Assuming a similar decay rate $\gamma$, the concurrences in Eqs.~\eqref{glob_conc} and \eqref{local_conc} can be immediately compared across models [see Fig.~\ref{fig:GMEdepolarizing}\,(a)]:
\begin{equation}
\gamma t_*^{\mathrm{glob}}\sim\frac{4}{3}\sin(2\theta),
\quad
\gamma t_*^{\mathrm{ind}}
\sim\left[\frac{\sqrt{2}}{3}\sin(2\theta)\right]^{2/3}.
\end{equation}
Except for the sufficiently weak initial entanglement, the local noise destroys the tripartite correlations earlier compared to the global white noise, as can be seen from Fig.~\ref{fig:GMEdepolarizing}\,(b).
Especially, for the maximally entangled GHZ state at $\theta = \pi/4$, we find:
\begin{equation}
\gamma t_*^{\rm ind} \simeq 0.335
\quad < \quad
\gamma t_*^{\rm glob} \simeq 0.847.
\end{equation}
However, this ordering is different for the initial states with $2\theta < \sin^{\!-1}(3/32)$ with $\gamma t_*^{\rm ind} > \gamma t_*^{\rm glob}$ as highlighted in the inset of Fig.~\ref{fig:GMEdepolarizing}\,(b).
For such cases, independent depolarization preserves genuine correlations longer.

This behavior can be understood by observing how the noise affects complementary basis states. 
For instance, beginning with state $|000\rangle$, applying independent noise for each qubit percolates to a state such as $|001\rangle$ through one flip, but reaching its complement $|110\rangle$ requires two flips.
In contrast, global depolarization affects all states at the same time.
Because the geometric mean of populations of complementary states enters the concurrence, these distinct population pathways can compensate for, and even outweigh, the faster decay of coherence under independent noise.
Thus, fixing the reduced dynamics of every qubit does not fix the ESD time of the multipartite state. The lifetime of genuine tripartite entanglement also depends on how the noise acts on the joint system.

\subsection{Extension to N-qubit GHZ state}
This construction can be directly extended to an $N$-qubit GHZ state.
Beginning with \begin{equation}
|\psi_0\rangle = \cos\theta\,\lvert{0}\rangle^{\otimes N} + \sin\theta\, \lvert{1}\rangle^{\otimes N},
\end{equation}
depolarization at the level of joint state has the following action:
\begin{equation}
\rho(0) \to \rho(t) = p(t) |\psi_0\rangle\langle \psi_0| + [1 - p(t)] \frac{\mathbb{I}_{2^N}}{2^N}.
\end{equation}
The concurrence for this $X$-state will be,
\begin{equation}
\label{eq:global_N_concurrence}
C_{\rm GME}^{\rm glob}(t, \theta) = \max\biggl\{0,\, p(t) \sin(2\theta) - [1-p(t)]\frac{2^N - 2}{2^N}\biggr\}.
\end{equation}
The genuine entanglement then completely depletes at
\begin{equation}
\label{eq:asymptotic_global}
t_*^{\rm glob} = \frac{1}{\gamma} \ln\biggl[1 + \frac{\sin (2\theta)}{1 - 2^{1 - N}}\biggr].
\end{equation}
Especially, the maximally entangled GHZ state $\theta = \pi/4$ loses genuine tripartite entanglement at
\begin{equation}
\label{eq:global_N_ESD_time}
\gamma t_*^{\rm glob}
=
\ln\left(
\frac{2^N-1}{2^{N-1}-1}
\right).
\end{equation}

On the other hand, for independent qubits evolving under single-qubit depolarizing map from Eq.~\eqref{eq:single_qubit_depol}, we find the final reduced state to be
\begin{equation}
\rho_{\rm loc}(t) = \mathcal{L}_p^{\otimes n}(|\psi_\theta\rangle \langle \psi_\theta|).
\end{equation}
Introducing $u = (1+p)/2$ and $v = (1-p)/2$,
each computational-basis state containing $k$ ones has population,
\begin{equation}
P_k = \cos^2(\theta) u^{N - k} v^k + \sin^2(\theta) v^{N-k} u^k.
\end{equation}
Its complimentary state then has population $P_{N - k}$.
The only non-zero off-diagonal entries are
\begin{equation}
\rho_{0\cdots0, 1\cdots1} = \rho_{1\cdots1, 0\cdots0} = p^N \cos(\theta) \sin(\theta).
\end{equation}
In total, there are $\binom{N}{k}$ basis states of weight $k$.
We arrive at the $X$-state concurrence after some simplifications,
\begin{align}
C_{\rm GME}^{\rm loc}(t, \theta) = \max\biggl\{
0,\
& p^N \sin(2\theta) \\
\notag & - \sum_{k = 1}^{N - 1}\binom{N}{k}\sqrt{P_k P_{N-k}}
\biggr\}.
\end{align}
One can verify that this expression reduces to Eq.~\eqref{local_conc} for $N = 3$.

Although the resulting threshold above does not admit a simple closed form for arbitrary $\theta$ and $N$, considerable simplification can be made for the maximally entangled GHZ state, $\theta=\pi/4$. In this case,
\begin{equation}
P_k=P_{N-k}
=
\frac{1}{2}
\left(
u^{N-k}v^k+v^{N-k}u^k
\right),
\end{equation}
so that the second part of non-zero contribution to the concurrence simplifies to 
\begin{align}
\sum_{k=1}^{N-1}
\binom{N}{k}\sqrt{P_kP_{N-k}}
&=
\sum_{k=1}^{N-1}\binom{N}{k}P_k
\nonumber\\
&=
1-u^N-v^N.
\end{align}
The concurrence therefore reduces to
\begin{equation}
\label{eq:CN_ind_GHZ}
C_{\rm GME}^{\rm loc}(t,\pi/4)
=
\max\bigl\{
0,\,
p^N+u^N+v^N-1
\bigr\}.
\end{equation}
The corresponding sudden-death point $p_*^{\rm loc}$ is determined implicitly by
\begin{equation}
\label{eq:pstar_ind_N}
\left(p_*^{\rm loc}\right)^N
+
\left(\frac{1+p_*^{\rm loc}}{2}\right)^N
+
\left(\frac{1-p_*^{\rm loc}}{2}\right)^N
=1,
\end{equation}
and consequently 
$\gamma t_*^{\rm loc} = -\ln p_*^{\rm loc}$, which is markedly different from that of global depolarization in Eq.~\eqref{eq:global_N_concurrence}.

The distinction between these two models is further evident in the large-$N$ limit. Specifically, for global depolarization channel acting on maximally entangled GHZ state, from Eqs.~\eqref{eq:global_N_concurrence} and~\eqref{eq:global_N_ESD_time},
\begin{equation}\label{eq:asymptotic_global:GHZ}
p_*^{\rm glob}\to\frac{1}{2},
\qquad
\gamma t_*^{\rm glob}\to\ln 2,
\end{equation}
so that the lifetime of genuine multipartite entanglement remains finite even as $N \to \infty$.

On contrary, local depolarization behaves qualitatively differently. 
To see this, notice that the large-\(N\) behavior of the local noise model follows directly from Eq.~\eqref{eq:pstar_ind_N}. Specifically, if $p_*^{\rm loc}$ is held fixed below unity as $N \to \infty$, then each term on the left-hand side of Eq.~\eqref{eq:pstar_ind_N} would vanish exponentially, contradicting the requirement that their sum must be equal to one. Therefore, we require $p_*^{\rm loc} \to 1$ for large $N$ which prompts us to consider the ansatz $p_*^{\rm loc} = 1 - \delta_N$ for $\delta_N \ll 1$. Then, the first two terms in Eq.~\eqref{eq:pstar_ind_N} become $(1-\delta_N)^N$ and $(1-\delta_N/2)^N$, while the last term vanishes in the large-$N$ limit. 
For these terms to remain finite and nontrivial in the large-$N$ limit, we need $N\delta_N=\mathcal{O}(1)$. 
Indeed, if $N\delta_N \to 0$, both terms approach unity, whereas if $N\delta_N\to\infty$, both vanish. 
We therefore set $\delta_N \sim a/N$, and the ansatz updates to
\begin{equation}
p_*^{\rm ind}
=
1-\frac{a}{N}
+\mathcal{O}(N^{-2}),
\end{equation}
with $a$ independent of $N$. 
Substitution into Eq.~\eqref{eq:pstar_ind_N} gives, to leading order,
\begin{equation}
e^{-a} + e^{-a/2} = 1,
\end{equation}
since the term $\bigl[(1-p_*^{\rm loc})/2\bigr]^N$ vanishes faster than exponentially. Thus, we obtain
$a = 2\ln\varphi$ with 
$\varphi=(1+\sqrt{5})/2$.
Consequently, we obtain for the ESD time
\begin{equation}
\label{eq:asymptotic_local}
\gamma t_*^{\rm loc}
\simeq
\frac{2\ln\varphi}{N} 
\simeq
\frac{0.96}{N}.
\end{equation}
In summary, for the maximally entangled GHZ state, using Eqs.~\eqref{eq:asymptotic_global:GHZ} and~\eqref{eq:asymptotic_local}, the asymptotic entanglement lifetimes therefore scale as,
\begin{equation}
t_*^{\rm loc}
=
\frac{2\ln\varphi}{\gamma N} + \mathcal{O}(N^{-2}),
\qquad
t_*^{\rm glob}
\longrightarrow
\frac{\ln 2}{\gamma}.
\end{equation}
Thus, the lifetime under local depolarization vanishes as $1/N$, whereas global depolarization retains a finite entanglement lifetime in the thermodynamic limit.
Interestingly, the asymptotic limit for local noise is independent of the initial state superposition $\theta$, whereas the global noise retains memory of this parameter as shown in Eq.~\eqref{eq:asymptotic_global}.

The different large-$N$ behaviors originate from how local and global depolarization modify the two ingredients entering the concurrence threshold. Under independent depolarization, the GHZ coherence between $\lvert0\rangle^{\otimes N}$ and $\lvert1\rangle^{\otimes N}$ is suppressed as $p^N$, while population is simultaneously transferred into an exponentially growing number of intermediate basis states. These populations increase the subtraction term in $C_{\rm GME}^{\rm loc}$, so increasing $N$ both accelerates the loss of coherence and strengthens the population contribution that destroys genuine multipartite entanglement. By contrast, under global depolarization, the GHZ coherence is reduced only by a single factor of $p$, while the added population is distributed uniformly over the full $2^N$-dimensional Hilbert space.

Even with the same reduced dynamics for every individual qubit, the genuine multipartite entanglement lifetime scales as $t_*^{\rm ind}\sim 1/N$ under independent depolarization, while $t_*^{\rm glob}$ approaches the finite value $\ln(2/\gamma)$ under global depolarization. Local dynamics alone therefore determines neither the ESD time nor its scaling with system size.

\section{Perspective and conclusions}

Eberly's work on entanglement sudden death firmly established that entanglement need not inherit the decay law of local coherence. Populations and coherences may evolve smoothly and remain nonzero while the entanglement of the composite system vanishes at a finite time. Studies across different noise models and physical settings showed that this behavior is not unique to spontaneous emission nor to a particular measure of entanglement. The lifetime of entanglement depends on the initial composite state and on how that state evolves under its environment. Eberly's explorations shifted focus from asking how fast quantum correlations decay to asking when, and under what conditions, a composite state loses its entanglement, and potentially even show signs of revival.

As the inquiry pivoted to multipartite systems, the subtleties grew in multiple folds. 
For two qubits, the disappearance of entanglement renders a separable state. 
In a multipartite system, vanishing of genuine multipartite entanglement can occur while entanglement remains across particular bipartitions or among smaller subsets of the system. 
Eberly's later work identified this distinction quantitatively. The exact genuine multipartite concurrence for $N$-qubit $X$-states allowed the finite-time loss of all-party entanglement to be followed directly. Furthermore, Eberly formulated genuine measures of entanglement for multipartite systems as well which later assisted the effort in discovering ESD for such systems.

The final two sections of this article return to these ideas directly. Section~III revisits the canonical bipartite mechanism of ESD, in which a smoothly evolving state reaches the separability boundary at a finite time. Section~IV gives, to the best of our knowledge, a new multipartite result. We study the spatial distribution of noisy evolution via global and local depolarizing dynamics that produce exactly the same reduced evolution for every individual qubit, while driving the full three-qubit state differently. 
Despite this identical local dynamics, the two processes give different death times for genuine tripartite entanglement. More strikingly, their ordering reverses for sufficiently weak initial entanglement. 
Thus, the lifetime of genuine multipartite entanglement is not fixed by the local dynamics: two noise processes that are indistinguishable at the single-qubit level can produce different ESD times, and even different large-system scaling. We believe this provides a direct continuation of multipartite entanglement decay that runs through Eberly's work.

What began with the question of whether entanglement should decay like coherence, developed through Eberly's work into a much broader study of the lifetime, transfer, revival, control, and multipartite structure of quantum entanglement. The discovery of ESD was an important part of that development, but its significance went beyond the identification of a new decay phenomenon. Eberly's work showed that the dynamics of entanglement must be understood at the level of the composite quantum system itself. That perspective, from the earliest studies of two-qubit ESD to his later work on genuine multipartite entanglement, remains a lasting part of his legacy in quantum physics.

\begin{acknowledgments}
A.S.H. thanks Gabriel Landi for helpful discussions, and acknowledges the funding support from the U.S. Department of Energy (DOE), Office of Science, Basic Energy Sciences (BES) under Award No. DE-SC0025516.
\end{acknowledgments}

\bibliography{references}

\end{document}